\documentclass[useAMS,usenatbib]{mnras}
\usepackage{times}
\usepackage{color}
\usepackage{color}
\usepackage{mathrsfs}
\usepackage{multicol}
\usepackage{bm}		
\usepackage{pdflscape}	
\usepackage{amsfonts,amsmath,amssymb,mathrsfs}
\usepackage{graphicx,epsf,epsfig}
\usepackage{inputenc}
\usepackage{bm}
\usepackage{longtable}
\usepackage[usenames,dvipsnames]{xcolor}
\usepackage{multirow}
\usepackage{hyperref}
\usepackage{times}
\usepackage{color}
\usepackage{tabularx}
\usepackage{appendix}

\def\be{\begin{equation}}
\def\ee{\end{equation}}
\def\bea{\begin{eqnarray}}
\def\eea{\end{eqnarray}}

\title[Testing DE Models with GRBs from OHD]{Testing Dark Energy Models with Gamma-Ray Bursts Calibrated from the Observational $H(z)$ Data through a Gaussian Process}
\author[Li, Zhang and Liang]
{Zihao Li\thanks{lizihao@gznu.edu.cn}, Bin Zhang\thanks{binzhang@gznu.edu.cn} and Nan Liang\thanks{Corresponding author: liangn@bnu.edu.cn}
\\
$^1$Key Laboratory of Information and Computing Science Guizhou Province, Guizhou Normal University, Guiyang, Guizhou 550025, China\\
$^2$Joint Center for FAST Sciences Guizhou Normal University Node, Guiyang, Guizhou 550025, China
}

\date{Accepted XXX. Received YYY; in original form ZZZ}

\pubyear{2022}

\begin{document}
\label{firstpage}
\pagerange{\pageref{firstpage}--\pageref{lastpage}}
\maketitle

\begin{abstract}
  We use a cosmology-independent method to calibrate gamma-ray burst (GRB) from the observational Hubble data (OHD)  with the cosmic chronometers method. By using Gaussian Process to reconstruct OHD, we calibrate the Amati relation ($E_{\rm p}$--$E_{\rm iso}$) to construct a  GRB Hubble diagram with the A118 data set, and constrain Dark Energy models in a flat space with the Markov Chain Monte Carlo numerical method.
  With the cosmology-independent GRBs at $1.4<z\leq8.2$ in the A118 data set and the Pantheon sample of type Ia supernovae (SNe Ia) at $0.01<z\leq2.3$, we obtained $\Omega_{\rm m}$ = $0.379^{+0.033}_{-0.024}$, $h$ = $0.701^{+0.0035}_{-0.0035}$,  $w$ = $-1.25^{+0.14}_{-0.12}$, $w_a$ = $-0.84^{+0.81}_{-0.38}$ for the flat Chevallier-Polarski-Linder model at the 1$\sigma$ confidence level. %which favor a possible DE evolution ($w_a\neq0$) at the 1-$\sigma$ confidence region.
  %We find that the $\Lambda$CDM model is favoured respect to the $w$CDM model and the CPL model with the selection criteria.
  We find no significant evidence supporting deviations from the standard $\Lambda$CDM model.
\end{abstract}


\section{Introduction}
%In practice, astrophysical objects often struggle to reach the state of a standard candle due to various disturbances and precision problems with observation instruments. Type Ia supernovae (SN Ia) is often used as a standard candle and the maximum redshift observed by SNe Ia is about $z\sim2.3$ \citep{Scolnic2018}. Therefore, to explore the cosmic evolution at the high-redshift region, observations of luminous objects at higher red-shift than SNe Ia are required.
Gamma-ray bursts (GRBs) are the most intense bursts of high-energy gamma rays from cosmological space in a short period of time. %which are proposed as supplementary tools to detect the expansion history of the universe.
At present, the maximum redshift of GRB can reach at $z\sim9$ \citep{Cucchiara2011}, while the maximum redshift observed for SNe Ia is about $z\sim2$ \citep{Scolnic2018}.  Therefore, GRBs can be used to probe the universe at high-redshift beyond SNe Ia.
The so-called Amati relation, which connects the spectral peak energy and the isotropic equivalent radiated energy (the $E_{\rm p}$-${E}_{\rm iso}$ correlation) of GRBs, has been proposed by \cite{Amati2002}.
Utilizing the Amati relation and/or other luminosity relations of GRB %which are connections between measurable properties of the instantaneous gamma-ray emission and the luminosity or energy,
\citep{Fenimore2000,Norris2000,Ghirlanda2004a,Yonetoku2004,Liang2005,Firmani2006,Dainotti2008,Yu2009,Tsutsui2009a,Izzo2015,Dainotti2020,Xu2021,Hu2021,Wang2022,Liu2022a}, the GRBs have been used as cosmic probe to study the evolutionary history of our Universe and the properties of dark energy \citep{Schaefer2003,Dai2004,Ghirlanda2004b,Firmani2005,Xu2005,Liang2006,Wang2006,Ghirlanda2006,Schaefer2007,Amati2008,Amati2019,Wang2008,Liang2008,Capozziello2008,Kodama2008,Tsutsui2009b,Cardone2009,Capozziello2010,Liang2010,Liang2011,Wei2010,WangDai2011,Gao2012,Liu2015,Lin2016,Wang2016,Demianski2017b,Khadka2020,Khadka2021,Luongo2020,Luongo2021,Muccino2021,Shirokov2020,Montiel2021,Demianski2021,Cao2022a,Cao2022b,Dainotti2022a,Liu2022b,Liang2022,Dainotti2022b,Dainotti2022c,Dainotti2023,Mu2023}.
For reviews of GRB luminosity relations and their applications in cosmology, see e.g. %\cite{Ghirlanda2006}, \cite{Schaefer2007},
\cite{Amati2013}, \cite{Dainotti2018}, and \cite{Moresco2022}.

Due to the lack of a low-redshift sample, a FIDUCIAL cosmological model has been assumed to calibrate the GRB luminosity relation in the early cosmological studies \citep{Dai2004}. When using the model-dependent GRB data to constrain cosmological models, the so-called circularity problem is encountered \citep{Ghirlanda2006}.
%Therefore, it is a huge challenge to avoid this circularity problem.
In order to avoid the circularity problem of GRB in cosmological applications, \cite{Liang2008} proposed a cosmological model-independent method to calibrate the luminosity relations of GRBs by using the SNe Ia data.
%It is clear that in any cosmology, objects at the same red-shift should have the same luminosity distance.
%If considering SNe Ia as a first-order standard candle,
The luminosity distances of the low-redshift GRB data to calibrate the GRB luminosity relations are  directly from SNe Ia at the same redshift. Following the cosmology-independent calibration method from SNe Ia by the interpolation method \citep{Liang2008} or by other similar approaches \citep{Kodama2008,Cardone2009,Capozziello2010,Gao2012,Liu2015,Izzo2015,Demianski2017a}, the derived GRB data at high redshift can be used to constrain cosmological models by using the standard Hubble diagram method \citep{Capozziello2008,Capozziello2009,Tsutsui2009b,Wei2009,Wei2010,Liang2010,Liang2011,DP2011}.

%Observational Hubble data is used to instead of SNe Ia data to calibrate the low red-shift GRBs data, though there are only 32 observational Hubble Data and SNe Ia data have 1048 \cite{Scolnic2018}.
On the other hand, the simultaneous fitting method has also been  proposed to alleviate the circularity problem, which constrain the coefficients of the relationship and the parameters of the cosmological model simultaneously \citep{Amati2008,Li2008}. However, a specific cosmological model is required for simultaneous fitting. %that the circularity problem cannot be avoided entirely by statistical methods.
Recently, \cite{Khadka2020} fitted the cosmological and GRB relation parameters simultaneously in six different cosmological models, and found that the Amati relation parameters are almost identical in all cosmological models.
\cite{Khadka2021} compiled a data set of 118 GRBs (the A118 sample) from the total 220 GRBs (the A220 sample) with the smallest intrinsic dispersion to simultaneously derive the correlation and cosmological model parameter constraints. \cite{Cao2022a,Cao2022b} used the A118 sample and the A220 sample to constrain cosmological model parameters simultaneously.

It should be noted that some systematic unknown biases of SN Ia
may propagate into the calibration results in the calibration precudure by using SNe Ia sample.
%Similar to the calibration method by using SNe Ia, the luminosity relations of GRB can be calibrated by using other observations at low redshift.
%The Hubble parameters have unique advantages in limiting cosmological parameters and distinguishing models.
%The observational Hubble data (OHD) obtained with the cosmic chronometers (CC) method, which related the evolution of differential ages of passive galaxies at different redshifts \citep{Jimenez2002}, have unique advantages to calibrated GRBs in a model-independent way.
%In order to remove these systematic biases of SN Ia,
\cite{Amati2019} proposed an alternative  method to calibrate GRB correlations by using the observational Hubble data (OHD)  at $z<1.975$ through the B\'ezier parametric curve and built up a Hubble diagram consisting of 193 GRBs.
Following this method, % from OHD fitted by the B\'ezier parametric to calibrate GRBs \citep{Amati2019},
\cite{Montiel2021} calibrate the Amati relation from the \emph{Fermi}-Gamma-ray Burst Monitor (GBM) catalogue (74 GRBs) with the OHD  at $z<1.43$ to approximate the cosmic expansion.
\cite{Luongo2021} used three machine learning treatments to get the mock data from differential Hubble rate points that calibrated the GRB correlations.
They proposed a model-independent calibration of GRB correlations in non-flat cosmology based on the most updated OHD and baryonic acoustic oscillations (BAO) \citep{Luongo2023}.
\cite{Muccino2022} calibrated the  Amati relation from 1000 simulated OHD points by means of the B\'ezier polynomial technique  to  constrain the transition redshift.

More recently,
%\cite{Wang2022} use a tight correlation between the plateau luminosity and the end time of the plateau in the X-ray afterglows out to the redshift $z = 5.91$. \cite{Jia2022} compiled a long GRB sample from Swift and Fermi observations, which contains 221 long GRBs with redshifts from 0.03 to 8.20.
\cite{Liu2022a} propose the improved Amati relations by accounting for evolutionary effects via a powerful statistical tool called copula, and  calibrated the copula relations from SNe Ia by the interpolation method to constrain cosmological models \citep{Liu2022b}.
\cite{Liang2022} calibrate the Amati relation with the A220 sample and the A118 sample by using a Gaussian process from SNe Ia and constrain on the $\Lambda$CDM model and the $w$CDM model in flat space with GRBs at high redshift and 31 OHD, which are consistent with those from fitting the coefficients of the Amati relation  and the cosmological parameters simultaneously.
%Since the absolute luminosity of SNe Ia data is unknown during calibration procedure, it should be fitted simultaneously with the calibration parameters.

In this paper, we calibrate the Amati relation  with the A118 GRB data \citep{Khadka2021} from the latest OHD by the Gaussian Process at low redshift, and  obtain the GRB Hubble diagram at  high redshift. Combining GRB data at $1.4 < z \leq 8.2$ with the Pantheon SNe Ia sample \citep{Scolnic2018}, we investigate Dark Energy (DE) models %(the $\Lambda$CDM model, the $w$CDM model, and the CPL model)
in a flat space with the Markov Chain Monte Carlo (MCMC) numerical method. %In addition, we also use the joint data sets to constrain DE models and GRB relation parameters simultaneously.
Finally, we compare the DE models with
the Akaike information criterion (AIC) and the Bayesian information criterion (BIC).

\section{CALIBRATION of Amati relation AT LOW-REDSHIFT}

%The Hubble parameters have unique advantages in limiting cosmological parameters and distinguishing models. So far, there are three methods to obtain the observation data of Hubble parameters: galactic age differential method, radial BAO size method and gravitational wave flute method.
%For OHD data set, we use the latest 32 OHD data set \citep{Zhang2022}. %Furthermore, we use python package emcee to implement Markov Chain Monte Carlo (MCMC) numerical method.

A Gaussian process is a non-parametric reconstruction for smoothing data by fully Bayesian approach, which has been widely applied to the field of cosmology \citep{Seikel2012a,Seikel2012b,Lin2018,Li2021,Benisty2021,Benisty2023}. In this work, we use the latest OHD for reconstruction by the Gaussian Process to calibrate Amati relation at low redshift.
Following \cite{Seikel2012a}, we use the squared exponential covariance function for the advantage that it is infinitely differentiable to reconstruct a function, which is given by
\begin{equation}
  k(z,\tilde{z})=\sigma_f^2\exp\left[-\frac{(z-\tilde{z})^2}{2l^2}\right].
\end{equation}
Here the hyperparameter $\sigma_f$ and $l$ can be optimized by maximizing the marginal likelihood. We use public PYTHON package GaPP \footnote{\url{https://github.com/astrobengaly/GaPP}} to reconstruct OHD at low redshift.

The model-independent OHD have unique advantages to calibrate GRBs in a model-independent way \citep{Amati2019}.
OHD can be obtained with the cosmic chronometers (CC) method, which relates the evolution of differential ages of passive galaxies at different redshifts without assuming any cosmological model \citep{Jimenez2002},
 \begin{equation}
 H(z)=-\frac{1}{1+z}\frac{dz}{dt}\,.
 \end{equation}
%Here $dt/dz$ can be taken to be the look-back time differential change with redshift \citep{Moresco2020}.
Applying the CC approach to the passively evolving galaxies from the luminous red galaxy (LRG) sample, 11 $H(z)$ data in the range $z\leq1.8$ were obtained \citep{Jimenez2003,Simon2005,Stern2010}.
Based on another analysis, \cite{Moresco2012,Moresco2015,Moresco2016} obtained 15
additional OHD in the range $0.179\leq z\leq1.975$.
With a full-spectrum fitting technique, \cite{Zhang2014} determined 4 additional estimates of OHD at $z<0.3$, and \cite{Ratsimbazafy2017} obtained one point at $z=0.47$, respectively.
These 31 OHD have been widely used for cosmological purposes \citep{Capozziello2018,Amati2019,Li2020,Montiel2021,Luongo2021,Luongo2023,Vagnozzi2021,Dhawan2021,Liu2022b,Liang2022}.
More recently, \cite{Borghi2022}  explored a new approach to obtain a new OHD at $z=0.75$.
\cite{Jiao2022} proposed a similar approach to obtain a new point at $z=0.80$. \footnote{Considering these two measurements are
not fully independent and their covariance is not clear, \cite{Zhang2022} only use the point \cite{Jiao2022},  which taking advantage of the $~1/\sqrt 2$ fraction of systematic uncertainty, with other 31 OHD to calibrate of HII Galaxies. In this work, we also use the 31 OHD and the point \cite{Jiao2022}. One could either use the data from \cite{Borghi2022} alternatively with other 31 OHD to investigate cosmology \citep{Cao2022,Kumar2022a,Kumar2022b,Muccino2022,Favale2023}.}
For reviews of recent progress of OHD, see e.g. \cite{Moresco2022}.
The latest OHD  obtained with the CC method %and the two points \citep{Borghi2022} and \citep{Jiao2022}
are summarized in Table 1.

\setlength{\tabcolsep}{4mm}{
\begin{table}
 \begin{center}{
  \caption{The latest OHD and their 1$\sigma$ errors obtained with the CC method. \label{Hubble data}}
 \begin{tabular}{|c|c|c|c|} \hline\hline
 \cline{1-4} $H(z)$& $z$& $\sigma_{H(z)}$ & REFERENCES \\ \hline
69  & 0.09  & 12 & \cite{Jimenez2003}\\
\hline
83  & 0.17  & 8 & \\
77  & 0.27  & 14 & \\
95  & 0.4  & 17 & \\
117  & 0.9  & 23 & \\
168  & 1.3  & 17 & \cite{Simon2005}\\
177  & 1.43  & 18 & \\
140  & 1.53  & 14 & \\
202  & 1.75  & 40 & \\
\hline
97  & 0.48  & 62 & \cite{Stern2010}\\
90  & 0.88  & 40 & \\
\hline
75  & 0.1791  & 4 & \\
75  & 0.1993  & 5 & \\
83  & 0.3519  & 14 & \\
104  & 0.5929  & 13 & \\
92  & 0.6797  & 8 & \cite{Moresco2012}\\
105  & 0.7812  & 12 & \\
125  & 0.8754  & 17 & \\
154  & 1.037  & 20 & \\
\hline
69  & 0.07  & 19.6 & \\
68.6  & 0.12  & 26.2 & \\
72.9  & 0.2  & 29.6 & \cite{Zhang2014}\\
88.8  & 0.28  & 36.6 & \\
\hline
160  & 1.363  & 33.6 & \cite{Moresco2015}\\
186.5  & 1.965  & 50.4 & \\
\hline
83  & 0.3802  & 13.5 & \\
77  & 0.4004  & 10.2 & \\
87.1  & 0.4247  & 11.2 & \cite{Moresco2016}\\
92.8  & 0.4497  & 12.9 & \\
80.9  & 0.4783  & 9 & \\
\hline
89  & 0.47  & 49.6 & \cite{Ratsimbazafy2017}\\
\hline
98.8  & 0.75  & 33.6 & \cite{Borghi2022}\\
\hline
113.1  & 0.80  & 28.5 & \cite{Jiao2022}\\
\hline\hline
 \end{tabular}}
 \end{center}
 \end{table}}

According to the analysis of \cite{Moresco2020} that OHD was carried out through simulations in the redshift range
$0 < z < 1.5$, \cite{Montiel2021} calibrated the Amati relation with OHD at $z<1.43$. %we limit our Hubble data to this range and instead of using the 31 measurements of $H(z)$ used in \citep{Capozziello2018}.
In order to compare with the previous analyses\citep{Liu2022b,Liang2022}, we used a subsample of Hubble parameter with a redshift cutoff at $z=1.4$, which including 28 points.
For GRB sample, we use the A118 GRB data set with the higher qualities appropriate for cosmological purposes in the A220 GRB data set \citep{Khadka2021}. There are 20 GRBs at low red-shift $z < 1.4$ in the A118 GRB sample.
The reconstructed results from the Gaussian process
with the 1$\sigma$ uncertainty from OHD are plotted in Figure 1.%and the higher-quality A118 GRB data set

 \begin{figure}
\centering
\includegraphics[width=230px]{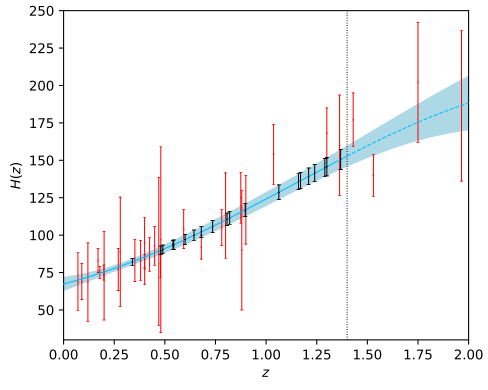}
\caption{The reconstructed results  from OHD through GaPP. The blue curves present the reconstructed function with the 1$\sigma$ uncertainty. The results of GRBs at $z<1.4$ (black dots) are reconstructed from OHD (red dots) through the Gaussian process. The black dotted line denotes $z=1.4$.} \label{fig/Hubble_GaPP.png}
\end{figure}

The Amati relation \citep{Amati2002}  connects the spectral peak energy ($E_{\rm p}$) and the isotropic equivalent radiated energy ($E_{\rm iso}$), which can be expressed as
\begin{equation}y = a + bx\end{equation}
where $y \equiv \log_{10}\frac{E_{\rm iso}}{1{\rm erg}},\quad x \equiv \log_{10}\frac{E_p}{300{\rm keV}}$, %$E_{iso}$ and $E_{p}$ are the isotropic equivalent radiated energy and the spectral peak energy respectively,
$a$ and $b$ are free coefficients needing to be calibrated from the GRBs observed data in the formula. $E_{\rm iso}$ and $E_{\rm p}$ can be respectively expressed as:
\begin{equation}E_{{\rm iso}} = 4\pi d^2_L(z)S_{{\rm bolo}}(1+z)^{-1},\quad E_p = E^{{\rm obs}}_p(1+z) \end{equation}
where $E^{\rm obs}_p$ is the observational value of GRB spectral peak energy and $S_{\rm bolo}$ is observational value of bolometric fluence, both $E^{{\rm obs}}_p$ and $S_{{\rm bolo}}$ can be observable.
The luminosity distance can be calculated by the reconstructed OHD at the redshift of GRBs\footnote{For the recent Planck results \citep{Plank2020} favor a flat spatial curvature, we consider a flat space in this work. However, recently works constrain non-spatially flat models with GRBs and results are promising \citep{Luongo2023}.}
\begin{eqnarray}\label{eqnarray4}
    d^{\rm GaPP}_{L} = c(1+z)\int^z_0\frac{dz^{'}}{H(z^{'})}
\end{eqnarray}
here the derivative $H(z)$ at redshift $z$ can be reconstructed with  OHD  by GaPP at the redshift of GRBs.

The parameters of Amati relation $a$ and $b$ can be fitted through GRBs sample data with $z < 1.4$ by using the likelihood function method \citep{D'Agostini2005}
\begin{eqnarray}\label{eqnarray5}
    \mathcal{L}_{\rm D}\propto\prod_{i=1}^{N_1} \frac{1}{\sigma}
    \times\exp\left[-\frac{[y_i-y(x_i,z_i; a, b)]^2}{2\sigma^2}\right]
\end{eqnarray}
Here $\sigma=\sqrt{\sigma_{\rm int}^2 + \sigma_{y,i}^2 + b^2\sigma_{x,i}^2}$, in which $\sigma_{\rm int}$ means the intrinsic scatter, $\sigma_{y} = \frac{1}{\ln{10}}\frac{\sigma_{E_{{\rm iso}}}}{E_{iso}}$, $\sigma_{E_{{\rm iso}}} = 4\pi d^{2}_{L}\sigma_{S_{{\rm bolo}}}(1+z)^{-1}$ is the error magnitude of isotropic equivalent radiated energy and $\sigma_{S_{{\rm bolo}}}$ means the error magnitude of bolometric fluence, $\sigma_{x} = \frac{1}{\ln{10}}\frac{\sigma_{E_{p}}}{E_{p}}$, $\sigma_{E_{p}}$ is the error magnitude of the spectral peak energy, $N_{1} =$ 20 means the number of low red-shift GRBs in A118 data set.
We implement Markov Chain Monte Carlo (MCMC) by using the PYTHON package EMCEE \citep{ForemanMackey2013}, which is optimized on the basis of Metropolis-Hastings algorithm. The calibrated results (the intercept $a$, the slope $b$ and the intrinsic scatter $\sigma_{\rm int}$) in the A118 GRB sample at $z < 1.4$ are summarized in Table 2.
We find that the results are consistent with previous analyses that obtained in \cite{Liang2022,Liu2022b} using GaPP and interpolation from SNe Ia at $z < 1.4$. %\textbf{by the likelihood method \citep{D'Agostini2005}}.
It should be noted that the use of the likelihood by \cite{D'Agostini2005} may introduce a subjective bias on the choice of the independent variable in the analysis. In order to get rid of this, \cite{Liang2008} used the same method (the bisector of the two ordinary least-squares; \cite{Isobe1990}) as used in \cite{Schaefer2007}.
\cite{Amati2013} used the likelihood function proposed by \cite{Reichart2001}, which has the advantage of not requiring the arbitrary choice of an independent variable among $E_{p}$ and $E_{{\rm iso}}$. The likelihood function proposed by \cite{Reichart2001} is written as \citep{Lin2016}
\begin{eqnarray}\label{eqnarray6}
    \mathcal{L}_{\rm R}\propto\prod_{i=1}^{N_1} \frac{\sqrt{1+b^2}}{\sigma}
    \times\exp\left[-\frac{[y_i-y(x_i,z_i; a, b)]^2}{2\sigma^2}\right]
\end{eqnarray}
Here $\sigma=\sqrt{\sigma_{\rm int}^2 + \sigma_{y,i}^2 + b^2\sigma_{x,i}^2}$, and the intrinsic scatter can be calculated by $\sigma_{\rm int}=\sqrt{\sigma_{y,\rm int}^2 + b^2\sigma_{x,\rm int}^2}$, in which $\sigma_{x,\rm int}$ and $\sigma_{y,\rm int}$ are the intrinsic scatter along the $x$-axis and $y$-axis.  The calibrated results in the A118 GRB sample at $z < 1.4$ are also summarized in Table 2. We find that the result of the slope $b$ by the likelihood method \citep{Reichart2001} is different with the one by the likelihood method \citep{D'Agostini2005}. In order to avoid any  bias on the choice of the independent variable, we decide to use the calibrated results by the likelihood method \citep{Reichart2001} to build the GRB Hubble diagram at $z>1.4$.

\setlength{\tabcolsep}{2mm}{
\begin{table}
 \begin{center}{
  \caption{Calibration results (the intercept $a$, the slope $b$, and the intrinsic scatter $\sigma_{\rm int}$) of the Amati relation in the A118 GRB sample at $z < 1.4$ by the likelihood method \citep{Reichart2001}  and the likelihood method \citep{D'Agostini2005}. \label{Amati result}}
 \begin{tabular}{|c|c|c|c|} \hline\hline
 \cline{1-4} Methods &$a$& $b$& $\sigma_{{\rm int}}$ \\ \hline
\cite{D'Agostini2005}&$52.87^{+0.11}_{-0.11}$  \ \ & \ \
$1.00^{+0.21}_{-0.21}$ \ \  & \ \
$0.49^{+0.06}_{-0.01}$\ \ \\
\cite{Reichart2001}&$52.84^{+0.15}_{-0.15}$  \ \ & \ \
$1.57^{+0.22}_{-0.42}$ \ \  & \ \
$0.55^{+0.59}_{-0.44}$\ \
\\
\hline\hline
 \end{tabular}}
 \end{center}
 \end{table}}

\section{GRB Hubble diagram and CONSTRAINTS ON DE MODELS}

Assuming the calibration results of the Amati relation at $z<1.4$ are valid at high redshift, we can derive the luminosity distances of GRBs at $z>1.4$ and build the GRB Hubble diagram, which is plotted in Figure 2. The uncertainty of GRB distance modulus with the Amati relation can be expressed as
\begin{eqnarray}\label{eqnarray6}
    \sigma^{2}_{\mu} = (\frac{5}{2}\sigma_{\log_{\frac{E_{{\rm iso}}}{1{\rm erg}}}})^{2} + (\frac{5}{2\ln10}\frac{\sigma_{S_{{\rm bolo}}}}{S_{{\rm bolo}}})^{2}
\end{eqnarray}
where
\begin{eqnarray}\label{eqnarray7}
    \sigma^{2}_{\log_{\frac{E_{iso}}{1erg}}} = \sigma^{2}_{int} + (\frac{b}{\ln10}\frac{\sigma_{E_{p}}}{E_{p}})^{2} + \sum \bigg (\frac{\partial_{y}(x;\theta_c)}{\partial \theta_i} \bigg)^2C_{ii}
\end{eqnarray}
Here $\theta_c = {\sigma_{{\rm int}}, a, b}$, and $C_{ii}$ means the diagonal element of the covariance matrix of these fitting coefficients.

 \begin{figure}
\centering
\includegraphics[width=250px,clip]{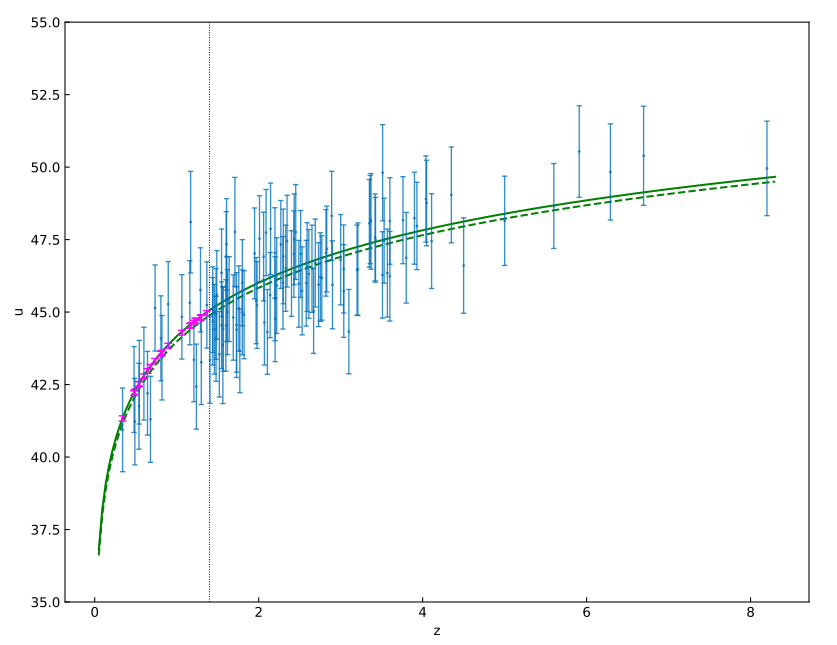}
\caption{GRB Hubble diagram with the A118 data set. GRBs at $z < 1.4$ were obtained by a Gaussian process from the OHD data (purple points), while GRBs with $z > 1.4$ (blue points) were obtained by the Amati relation and calibrated with A118 at $z < 1.4$. The solid green curve is the CMB standard distance modulus with $H_0$ = 67.36 km $s^{-1}$ ${{\rm Mpc}}^{-1}$, $\Omega_m$ = 0.315 \citep{Plank2020}, and the green long dotted curve is the SNIa standard distance modulus with $H_0$ = 74.3 km $s^{-1}$ ${{\rm Mpc}}^{-1}$, $\Omega_m$ = 0.298 \citep{Scolnic2018}. The black dotted line denotes $z = 1.4$.} \label{fig/GRB_Hubble.png}
\end{figure}

We use the GRB data in the Hubble diagram at $z>1.4$ with the Pantheon sample \citep{Scolnic2018} to constrain cosmological models.
The $\chi^2$ for the distance modulus can be expressed as
\begin{equation}\chi^2_{\mu} = \sum^{N}_{i=1} \left[\frac{\mu_{\rm obs}(z_i)-\mu_{\rm th}(z_i;p,H_0)}{\sigma_{\mu_i}}\right]^2.
\end{equation}
Here $\mu_{{\rm obs}}$ is the observational value of distance modulus and its error $\sigma_{\mu_i}$, and $\mu_{{\rm th}}$ is the theoretical value of distance modulus calculated from the cosmological model, $p$ represents the cosmological parameters.
The theoretical distance modulus of DE modles can be calculated as
\begin{eqnarray}\label{mu}
\mu=5\log \frac{d_L}{\textrm{Mpc}} + 25=5\log_{10}D_L-\mu_0,
\end{eqnarray}
where $\mu_0=5\log_{10}h+42.38$, $h=H_0/(100{\rm km/s/Mpc})$, $H_0$ is the Hubble constant.
For a flat space, the unanchored luminosity
distance $D_L$ can be calculated by
\begin{eqnarray}
D_L\equiv
H_0d_L=(1+z)\int_0^z\frac{dz'}{E(z')},
\end{eqnarray}
where $E(z)=[\Omega_{\textrm{M}}(1+z)^3+\Omega_{\textrm{DE}}X(z)]^{1/2}$, and  $X(z)=\exp[3\int_0^z\frac{1+w(z')}{1+z'}dz']$, which is determined by the choice of the specific dark energy (DE) model.
We consider three DE models in a flat space, the $\Lambda$CDM model with dark energy EoS $w=-1$ , the $w$CDM model  with a constant Equation of State(EoS),  and the Chevallier-Polarski-Linder (CPL) model \citep{CP2001,Linder2003}  in which dark energy evolving with redshife as a parametrization  EoS,  $w=w_0+w_az/(1+z)$.
\begin{eqnarray}\label{eqnarrayCPL}
X(z)=\begin{cases}
1, & \rm{\Lambda CDM} \\
(1+z)^{3(1+w_0)},& w\rm{CDM} \\
(1+z)^{3(1+w_0+w_a)}e^{-\frac{3w_az}{1+z}}, & \rm{CPL} \\
\end{cases}\end{eqnarray}

The Pantheon sample contains 1048 SNe spanning the redshift range $0.01<z<2.3$, with
the observed distance modulus of SNe given by \citep{Scolnic2018}
\begin{equation}
    \mu_{\rm SN} = m_{\rm B}^* - M + \alpha X_1 - \beta \mathcal{C} + \Delta_M + \Delta_B,
    \label{eq:muSN}
\end{equation}
where $m_{\rm B}^*$ is the observed peak magnitude in rest frame B-band, $M$ is the absolute magnitude, $X_1$ is the time stretching of the light-curve, $\mathcal{C}$ is the SNe color at maximum brightness; $\alpha, \beta$  are nuisance parameters which should be fitted simultaneously with the cosmological parameters,
$\Delta_M$  is a distance correction
based on the host galaxy mass of the SN, and $\Delta_B$  is a distance
correction based on predicted biases from simulations. The Pantheon data set is calibrated using the Bayesian Estimation Applied to Multiple Species (BEAMS) with Bias Corrections (BBC) method \citep{Kessler2017}, and the corrected apparent magnitude for all the SNe have been reported in \cite{Scolnic2018}.
\begin{equation}
    m_{\rm B,corr}^* = m_{\rm B}^* + \alpha X_1 -\beta \mathcal{C}+ \Delta_M + \Delta_B,
\end{equation}
 %The $\chi^2_{SN}$ for the Pantheon data can be calculated by
%\begin{align}
%    \chi_{SN}^2={\bf \Delta \hat{\mu}^T} \cdot \textbf{Cov}^{-1} \cdot {\bf \Delta \hat{\mu}}
%\end{align}
%where $\Delta \hat{\mu}_i = m_{{\rm B,corr},i}^* - 5\log_{10}[d_L(z_i)] +(M_{\rm B}^* +\mu_0)$, and $M_{\rm B}^* + \mu_0$ can be marginalized over analytucally \citep{Lin2018,Li2020}.
Following \cite{Liu2022b}, we set the absolute magnitude to be $M=-19.36$ \citep{Gomez-Valent2022} in our analysis to obtain the distance modulus of SNe Ia.

\setlength{\tabcolsep}{0.3em}{
\begin{table*}
% \begin{table}[tbhp]
 \begin{center}{%\scriptsize
  \caption{Joint constraints on parameters of $\Omega_m$, $h$, $w_0$ and $w_a$ for the flat $\Lambda$CDM model, the flat $w$CDM model, and the CPL model with 98 GRBs ($z> 1.4$) + 1048 SNe. } \label{Joint constrain results}
 \begin{tabular}{|c|c|c|c|c|c|c|c|c|c|} \hline
 \cline{1-9}Models & Data Set & $\Omega_{m}$ & $h$ & $w_0$ & $w_a$  & $-2\ln \mathcal{L_{\rm R}}$  & $\Delta \rm AIC$  & $\Delta \rm BIC$  \\ \hline
$\Lambda$CDM                \ \ & \ \
98 GRBs  \ \  & \ \
$0.40^{+0.13}_{-0.34}$ \ \ & \ \
$0.72^{+0.11}_{-0.20}$ \ \ & \ \
-\ \ &\ \
-\ \ &\ \
43.828\ \ &\ \
-\ \ &\ \
-\ \ \\
$w$CDM                \ \ & \ \
98 GRBs  \ \  & \ \
$0.34^{+0.10}_{-0.33}$ \ \ & \ \
$0.712^{+0.082}_{-0.19}$ \ \ & \ \
$-0.97^{+0.70}_{-0.52}$\ \ &\ \
-\ \ &\ \
42.803\ \ &\ \
0.974\ \ &\ \
3.559\ \ \\
CPL                \ \ & \ \
98 GRBs  \ \  & \ \
$0.42^{+0.16}_{-0.38}$ \ \ & \ \
$0.695^{+0.088}_{-0.19}$ \ \ & \ \
$-0.90^{+0.86}_{-0.35}$\ \ &\ \
$-0.99^{+0.58}_{-0.58}$\ \ &\ \
46.173\ \ &\ \
6.344\ \ &\ \
11.514\ \ \\\hline
$\Lambda$CDM                \ \ & \ \
98 GRBs +  1048SNe \ \  & \ \
$0.286^{+0.012}_{-0.012}$ \ \ & \ \
$0.6970^{+0.0022}_{-0.0022}$ \ \ & \ \
-\ \ &\ \
-\ \ &\ \
1079.033\ \ &\ \
-\ \ &\ \
-\ \ \\
$w$CDM                \ \ & \ \
98 GRBs +  1048SNe \ \  & \ \
$0.350^{+0.036}_{-0.028}$ \ \ & \ \
$0.7019^{+0.0035}_{-0.0035}$ \ \ & \ \
$-1.25^{+0.15}_{-0.13}$\ \ &\ \
-\ \ &\ \
1075.932\ \ &\ \
1.102\ \ &\ \
3.943\ \ \\
CPL                \ \ & \ \
98 GRBs +  1048SNe \ \  & \ \
$0.379^{+0.033}_{-0.024}$ \ \ & \ \
$0.7010^{+0.0035}_{-0.0035}$ \ \ & \ \
$-1.25^{+0.14}_{-0.12}$\ \ &\ \
$-0.84^{+0.81}_{-0.38}$\ \ &\ \
1076.335\ \ &\ \
1.301\ \ &\ \
11.389\ \ \\
\hline\hline
 \end{tabular}}
 \end{center}
 \end{table*}}

The cosmological parameters can be fitted by using the minimization $\chi^2$ method through MCMC method. The total $\chi^2$ with the joint data of GRB+SNe can be expressed as
$\chi^2_{{\rm total}} = \chi^2_{{\rm GRB}} + \chi^2_{{\rm SN}}.$
The python package emcee \citep{ForemanMackey2013} is used to constrain DE models.
Constraints only with 98 GRBs (A118) $z > 1.4$ are shown in Figure 3 ($\Lambda$CDM) , Figure 4 ($w$CDM) and Figure 5 (CPL), which are summarized in Table 3. With 98 GRBs at $1.4<z\le8.2$, we obtain $\Omega_{\rm m}$ = $0.40^{+0.13}_{-0.34}$, $h$ = $0.72^{+0.11}_{-0.20}$ for the flat $\Lambda$CDM model, $\Omega_{\rm m}$ = $0.34^{+0.10}_{-0.33}$, $h$ = $0.712^{+0.082}_{-0.19}$  $w$ = $-0.97^{+0.70}_{-0.52}$ for the flat $w$CDM model at the 1$\sigma$ confidence level.
These results for the flat $\Lambda$CDM  and the flat $w$CDM model are consistent with previous analyses that obtained in \cite{Liang2022} using GaPP from SNe Ia at $z < 1.4$ by setting $H_0$=$70\ {\rm km}\ {\rm s}^{-1}{\rm Mpc}^{-1}$ for the cases only with GRBs.
The joint results from 98 GRBs (A118) $z > 1.4$ with 1048 SNe Ia  are shown in Figure 6 ($\Lambda$CDM), Figure 7 ($w$CDM) and Figure 8 (CPL), which are summarized in Table 3.
With 98 GRBs at $1.4<z<8.2$ in the A118 sample and 1048 SNe Ia, we obtained $\Omega_{\rm m}$ = $0.286^{+0.012}_{-0.012}$ and $h$ = $0.6970^{+0.0022}_{-0.0022}$  for the flat $\Lambda$CDM model, and  $\Omega_{\rm m}$ = $0.350^{+0.036}_{-0.028}$, $h$ = $0.7019^{+0.0035}_{-0.0035}$, $w$ = $-1.25^{+0.15}_{-0.13}$ for the flat $w$CDM model,
which are consistent with previous analyses that obtained in \cite{Liang2022} using GaPP from SNe Ia at $z < 1.4$.

For the flat CPL model with 98 GRBs $z > 1.4$,  we obtained $\Omega_{\rm m}$ = $0.42^{+0.16}_{-0.38}$, $h$ = $0.695^{+0.088}_{-0.19}$,  $w$ = $-0.90^{+0.86}_{-0.35}$, $w_a$ = $-0.99^{+0.58}_{-0.58}$;
with 98 GRBs $z > 1.4$ and 1048 SNe Ia, we obtained $\Omega_{\rm m}$ = $0.379^{+0.033}_{-0.024}$, $h$ = $0.7010^{+0.0035}_{-0.0035}$,  $w$ = $-1.25^{+0.14}_{-0.12}$, $w_a$ = $-0.84^{+0.81}_{-0.38}$  for the flat CPL model at the 1$\sigma$ confidence level, which favor a possible DE evolution ($w_a\neq0$) at the 1-$\sigma$ confidence region for both cases.
%which indicate that the $\Lambda$CDM model ($w=-1$) is not not consistent with the joint data at the 1-$\sigma$ confidence region.

\begin{figure}
\centering
\includegraphics[width=\hsize,clip]{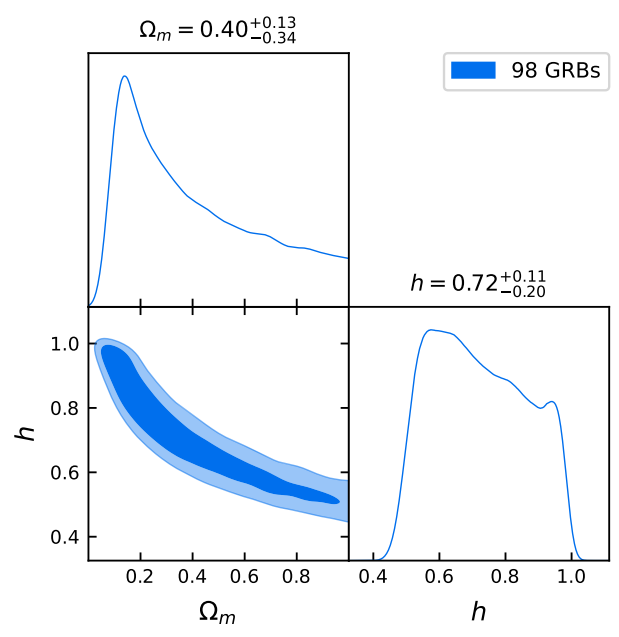}
\caption{Constraints on parameters of $\Omega_m$, $h$ for the flat $\Lambda$CDM model  with 98 GRBs at $z > 1.4$.}
\end{figure}

\begin{figure}
\centering
\includegraphics[width=\hsize,clip]{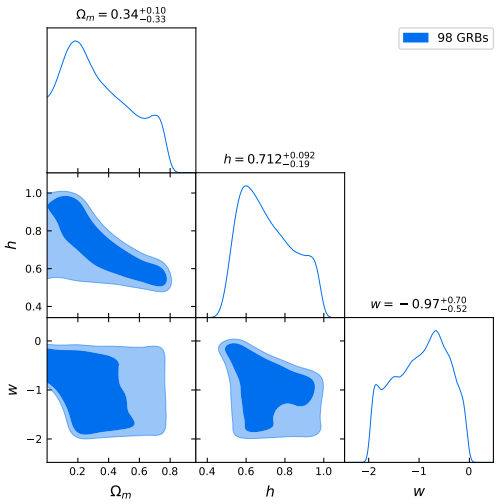}
\caption{Constraints on parameters of $\Omega_m$, $h$, and $w_0$  for the flat $w$CDM model with 98 GRBs at $z > 1.4$.}
\end{figure}

\begin{figure}
\centering
\includegraphics[width=\hsize,clip]{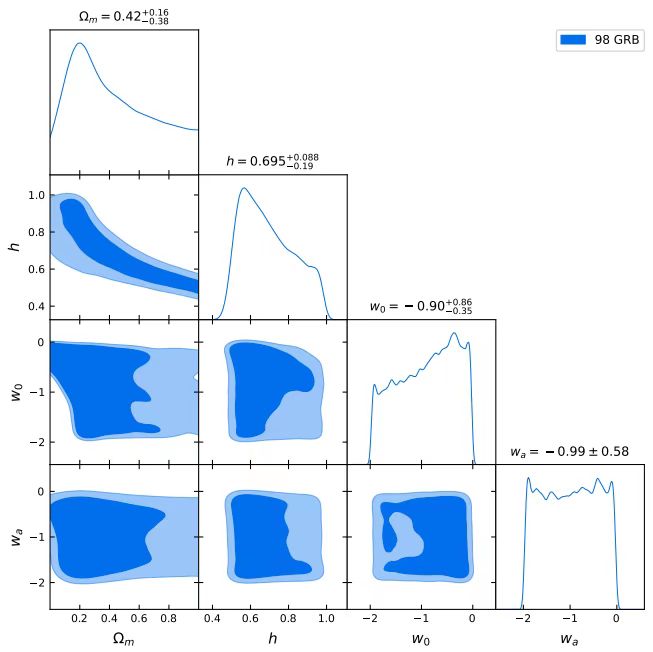}
\caption{Constraints on parameters of $\Omega_m$, $h$, $w_0$ and $w_a$  for the flat CPL model  with 98 GRBs at $z > 1.4$.}
\end{figure}

\begin{figure}
\centering
\includegraphics[width=\hsize,clip]{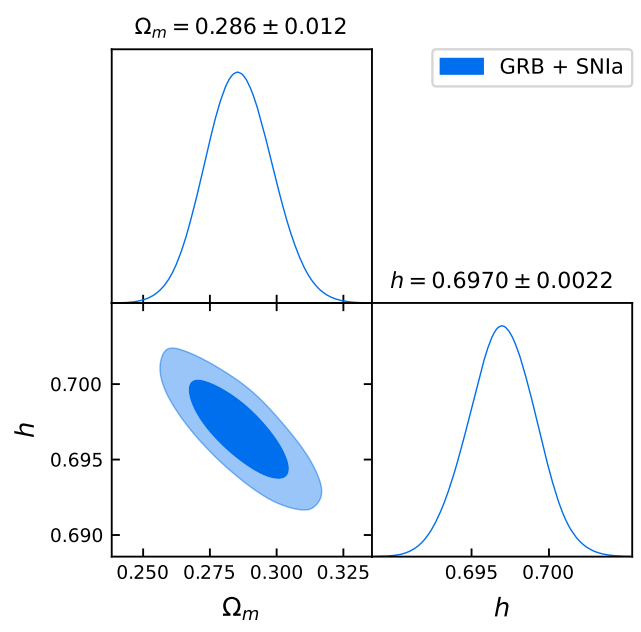}
\caption{Joint constraints on parameters of $\Omega_m$, $h$ for the flat $\Lambda$CDM model with 98 GRBs ($z > 1.4$ ) + 1048 SNe.}\label{Hubble_LambdaCDM}
\end{figure}

\begin{figure}
\centering
\includegraphics[width=\hsize,clip]{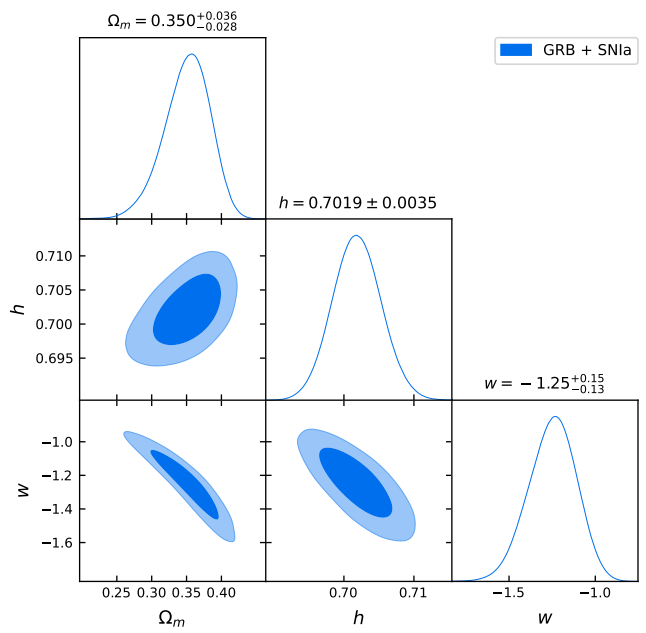}
\caption{Joint constraints on parameters of $\Omega_m$, $h$, and $w_0$  for the flat $w$CDM model  with 98 GRBs ($z > 1.4$ ) + 1048 SNe.} \label{Hubble_wCDM}
\end{figure}

\begin{figure}
\centering
\includegraphics[width=\hsize,clip]{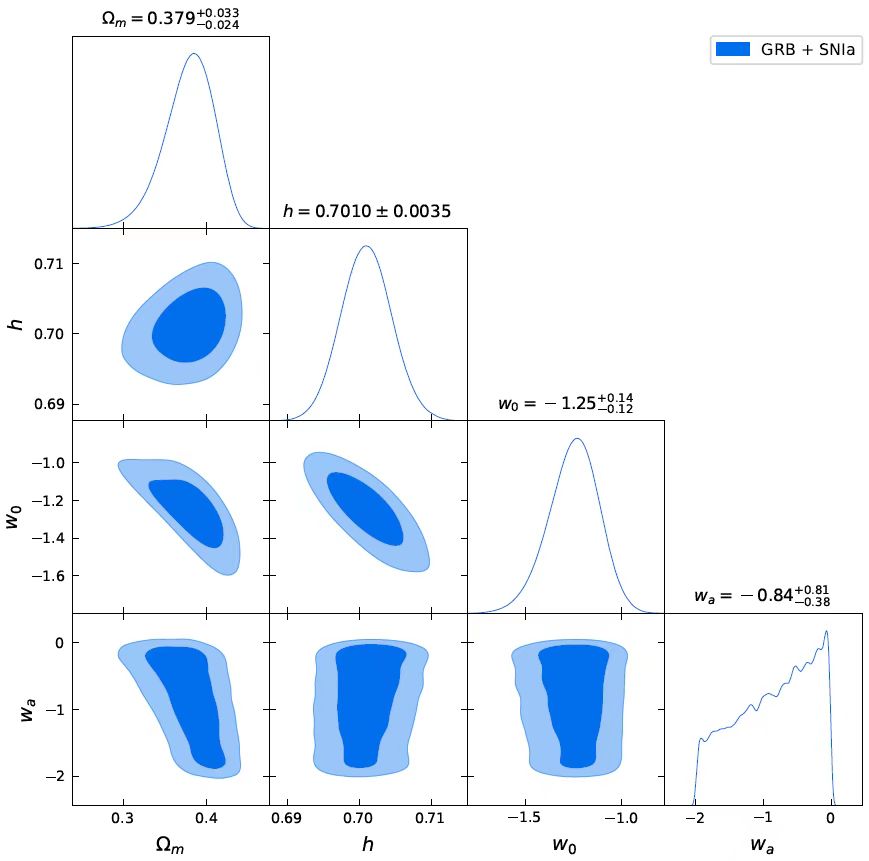}
\caption{Joint constraints on parameters of $\Omega_m$, $h$,  $w_0$ and $w_a$ for the flat CPL model with 98 GRBs ($z> 1.4$ ) + 1048 SNe.} \label{CPL}
\end{figure}

In order to compare the different cosmological models, we compute the values of the Akaike information criterion (AIC; \cite{Akaike1974,Akaike1981}) and the Bayesian information criterion (BIC; \cite{Schwarz1978}), respectively.
\begin{eqnarray}\label{eqnarray_AIC} \rm AIC = 2\emph{p}-2\ln(\mathcal{L})\end{eqnarray}
\begin{eqnarray}\label{eqnarray_BIC} \rm BIC = \emph{p}\ln \emph{N}-2\ln(\mathcal{L})\end{eqnarray}
where $\mathcal{L}$ is the maximum value of the likelihood function, $ p $ is the number of free parameters in a model, and $ N $ is the number of data. The value of $\Delta \rm AIC$ and $\Delta \rm BIC$, which denotes the difference between $\rm AIC$ and $\rm BIC$  with respect to the reference  model (the $\Lambda$CDM model) are summarized in Table 3.
For the value of $\Delta \rm AIC$ and $\Delta \rm BIC$,  $0 < \Delta \rm AIC(\Delta \rm BIC) < 2$ indicates difficulty in preferring a given model, $2 < \Delta \rm AIC(\Delta \rm BIC) < 6$ means mild evidence against the given model, and $\Delta \rm AIC(\Delta \rm BIC) > 6$ suggests strong evidence against the model.
We find that the results of $\Delta \rm AIC$ and $\Delta \rm BIC$ indicate that the $\Lambda$CDM model is favoured respect to the $w$CDM model and the CPL model, which are consistent with the previous analyses \citep{Amati2019} obtained from the 193 GRBs by using the OHD  at $z<1.975$ through the B\'ezier parametric curve combined with 740 SNe Ia.

\section{CONCLUSIONS AND DISCUSSIONS}
In this paper, we use the Gaussian process to calibrate the Amati relation from OHD and obtain the GRB Hubble diagram with the A118 sample. With 98 GRBs at $1.4<z<8.2$ in the A118 sample and 1048 SNe Ia, we obtained $\Omega_{\rm m}$ = $0.286^{+0.012}_{-0.012}$ and $h$ = $0.6970^{+0.0022}_{-0.0022}$  for the flat $\Lambda$CDM model, and  $\Omega_{\rm m}$ = $0.350^{+0.036}_{-0.028}$, $h$ = $0.7019^{+0.0035}_{-0.0035}$, $w$ = $-1.25^{+0.15}_{-0.13}$ for the flat $w$CDM model,
and $\Omega_{\rm m}$ = $0.379^{+0.033}_{-0.024}$, $h$ = $0.7010^{+0.0035}_{-0.0035}$,  $w$ = $-1.25^{+0.14}_{-0.12}$, $w_a$ = $-0.84^{+0.81}_{-0.38}$  for the flat CPL model at the 1$\sigma$ confidence level, which favor a possible DE evolution ($w_a\neq0$) at the 1-$\sigma$ confidence region for both cases. We find that the results of $\Delta \rm AIC$ and $\Delta \rm BIC$ indicate that the $\Lambda$CDM model is favoured respect to the $w$CDM model and the CPL model.
In order to compare with simultaneous fitting method, we also use GRB data sets of A118 sample and SNe Ia to fit the coefficients of the Amati relation ($a$, $b$, $\sigma_{\rm int}$) and the cosmological parameters  ($\Omega_{\rm m}$, $h$, $w$, and $w_a$) simultaneously for the flat $\Lambda$CDM model, the flat $w$CDM model and the flat CPL model. It is found that the simultaneous fitting results are consistent with those obtained from the low-redshift calibration method.

It should be notice that GRB luminosity relations can be calibrated by using other observations, besides the calibration method by using SN Ia and OHD. For examples,  \cite{Wang2019} used the mock gravitational waves (GWs) catalog as standard sirens to calibrate GRB luminosity correlations.
\cite{Dai2021} calibrated GRBs from quasar sample at $0.5 < z < 5.5$ which divided into several subsamples with different redshift bins, and found that the Amati relation shows no evolution with redshift.
\cite{Gowri2022} used the angular diameter distances of 38 galaxy clusters to circumvent the circularity problem in the Amati relation. Moreover, whether the GRB relations are redshift dependent or not is still under debate \citep{Khadka2021,Dai2021,Tang2021,Liu2022a}.
Recently,
\cite{Wang2022} use a tight correlation between the plateau luminosity and the end time of the plateau in the X-ray afterglows out to the redshift $z = 5.91$. \cite{Jia2022} compiled a long GRB sample from Swift and Fermi observations, which contains 221 long GRBs with redshifts from 0.03 to 8.20.
Along with the GRB sample from Konus-Wind (KW), Swift, and GRBs observed from Fermi with much
smaller scatters, as well as the Chinese-French mission SVOM (the Space-based multiband astronomical Variable Objects Monitor), to be launched this year, which will provide a substantial enhancement of the number of GRBs with measured redshift and spectral parameters \citep{Bernardini2021},  GRBs could be used as an additional choice to
set tighter constraints on cosmological parameters of DE models.

\section*{ACKNOWLEDGMENTS}
We thank Prof. Puxun Wu, Prof. Xiaolei Li, Yang Liu, Hanbei Xie, XiaoDong Nong, Huifeng Wang, Guangzhen Wang, Zhiguo Xiong, and Prof. Xiaoyao Xie, Prof.  Jianchao Feng, Prof.  Junjin Peng  for kind help and discussions. We also thank the referee for  helpful comments and constructive suggestions.
This project was supported by the Guizhou Provincail Science and Technology Foundation (QKHJC-ZK[2021] Key 020).

\section*{DATA AVAILABILITY}
Data are available at the following references:
the latest OHD obtained with the CC method from
Table 1 and references therein, the A118 sample of GRB data set
from \cite{Khadka2021} and the Pantheon SN sample from \cite{Scolnic2018}.
The data underlying this article will be shared on reasonable request
to the corresponding author.

%\bibliographystyle{mnras}
%\bibliogra  phy{ref} % if your bibtex file is called %example.bib

%%%%%%%%%%%%%%%%%%%%%%%%%%%%%%%%%%%%%%%%%%%%%%%%%%

% Don't change these lines
\bsp	% typesetting comment
\label{lastpage}
\end{document}